# All-metallic Vertical Transistors Based on Stacked Dirac Materials

Yangyang Wang,[1] Zeyuan Ni,[1] Qihang Liu,[4] Ruge Quhe,[1,5] Jiaxin Zheng,[6] Meng Ye,[1] Dapeng Yu,[1] Junjie Shi,[1] Jinbo Yang,[1,2] Ju Li,[3] and Jing Lu[1,2*]

[1]State Key Laboratory for Mesoscopic Physics and Department of Physics, Peking University, Beijing 100871, P. R. China

[2]Collaborative Innovation Center of Quantum Matter, Beijing 100871, P. R. China

[3]Department of Nuclear Science and Engineering and Department of Materials Science and Engineering, Massachusetts Institute of Technology, Cambridge, Massachusetts 02139, USA

[4]University of Colorado, Boulder, Colorado 80309, USA

[5]Academy for Advanced Interdisciplinary Studies, Peking University, Beijing 100871, P. R. China

[6]School of Advanced Materials, Peking University, Shenzhen Graduate School, Shenzhen 518055, P. R. China

Corresponding author: jinglu@pku.edu.cn

**ABSTRACT**

It is a persisting pursuit to use metal as a channel material in a field effect transistor. All metallic transistor can be fabricated from pristine semimetallic Dirac materials (such as graphene, silicene, and germanene), but the on/off current ratio is very low. In a vertical heterostructure composed by two Dirac materials, the Dirac cones of the two materials survive the weak interlayer van der Waals interaction based on density functional theory method, and electron transport from the Dirac cone of one material to the one of the other material is therefore forbidden without assistance of phonon because of momentum mismatch. First-principles quantum transport simulations of the all-metallic vertical Dirac material heterostructure devices confirm the existence of a transport gap of over 0.4 eV, accompanied by a switching ratio of over $10^4$. Such a striking behavior is robust against the relative rotation between the two Dirac materials and can be extended to twisted bilayer graphene. Therefore, all-metallic junction can be a semiconductor and novel avenue is opened up for Dirac material vertical structures in high-performance devices without opening their band gaps.

## 1. Introduction

Since the semiconductor industry based on Si is approaching limit of performance improvement, it is a persisting pursuit to use metal as channel material in a field effect transistor (FET). All-metallic FETs could be scaled down to smaller size with less energy consume and performance at higher frequency.[1] No metal or semimetal has shown any notable field effect until the appearance of graphene.[2] Graphene is semimetal but its current is sensitive to electrical field due to its extreme thickness. However, the on/off current ratio of graphene is less than 30, and this greatly limits the application of pure graphene in electronics though its high carrier mobility of up to $10^5$ cm$^2$/(V·s) is very attractive.[2] After the discovery of graphene, silicene has been fabricated via epitaxial growth on the Ag(111),[3-5] ZrB$_2$(0001),[6] Ir(111),[7] and MoS$_2$ surfaces,[8] and germanene has also been grown on Pt (111) surfaces recently.[9] High carrier mobility is also calculated in silicene and can be expected in germanene,[10] but they suffer from the same obstacle—very poor on/off current ratio due to their zero band gaps.[11] A fundamental and intriguing question arises: Is it possible to fabricate a high performance FET with metal or semimetal?

Opening a band gap without degrading the mobility can pave the avenue for the application of a Dirac material in high-speed nanoelectronics. The existing approaches of meeting this requirement such as imposing a vertical electric field often suffer from a too small band gap (< 0.3 eV) and thus a poor on/off current ratio (< 1000), and this problem is especially prominent for graphene.[11-18] Any successful successor to the silicon metal-oxide-semiconductor FET (MOSFET) that is to be used in complementary MOS-like logic must have an on/off ratio of between $10^4$ and $10^7$, which requires a semiconducting channel with a transport gap of over 0.4 eV.[19] Therefore, it is highly desired to develop a new method to realize high-performance Dirac material FET devices with a transport gap of over 0.4 eV and an on/off current ratio of no less than $10^4$.

Stacking different two-dimensional (2D) atomic crystals provides a unique opportunity to create new layered materials. The properties of artificially stacked layered materials depend on the composition of 2D crystals and the stacking style, and thus have substantial tunability. So far, graphene/h-BN,[20-22] graphene/MoS$_2$,[22-26] and graphene/WS$_2$ [27] heterostructures have been successfully fabricated and serve as tunneling FETs with a high on/off current ratio up to $10^6$.[22, 24, 26, 27] Recently a Dirac material vertical heterostructure (graphene/silicene) has been grown on Ir(111) and Rh(0001) surfaces.[28, 29]

In this article, we reveal that the Dirac cones of the two Dirac materials in a vertical heterostructure survive the weak interlayer van der Waals interaction and are completely free from band hybridization near the Fermi level ($E_f$), suggesting that electron transport from one Dirac material to the other near $E_f$ is forbidden without assistance of phonon because of momentum mismatch. Although, this heterostructure is all-metallic, a large transport gap of over 0.4 eV is observed in an *ab initio* quantum transport simulation of a single-gated two-probe model, accompanied by a high on/off current ratio of over $10^4$. Such an intriguing property in Dirac material heterostructures is robust against the relative rotation of the two Dirac materials and can also be expanded to homogenous twisted bilayer graphene (BLG).

## 2. Results and Discussion
### 2.1. Geometry and Stability of (3×3)Graphene/(2×2)Silicene Heterostructures

Since adjacent layers are covalently bonded in multilayer silicene,[30-32] here only single layer silicene is used to combine with SLG (single layer graphene), BLG, and TLG (trilayer graphene). We first consider the matching patterns without relative rotations. A supercell model is constructed from a (3×3) graphene unit cell and a (2×2) silicene unit cell. We fix the in-plane lattice constant of the supercell to $a_S = 7.5$ Å, and the lattice constant deviations from the experimental values of graphene and silicene are 1.6% and 3%, respectively. Under this lattice mismatch, the characters of linear dispersion near $E_f$ are intact in both standalone graphene and silicene and the work functions of them only increase by 0.1 and 0.05 eV compared with the unstrained ones, respectively.

To find an energetically stable superlattice, three stacking patterns (I-III) of SLG/silicene heterostructures are considered, as shown in Figure 1. After relaxation, the top views of the three patterns keep unchanged. However, in Pattern I and II, three types of Si atoms in height denoted as $Si_A$, $Si_B$, and $Si_C$ are found, while in Pattern III, there only exist two types $Si_A$ and $Si_B$. The binding energy $E_b$ of graphene/silicene heterostructures between graphene and silicene is defined as

$$E_b = (E_{Gr/Si} - E_{Gr} - E_{Si})/N_C$$

Where $E_{Gr}$, $E_{Si}$, and $E_{Gr/Si}$ are the relaxed energies for graphene, silicene, and the combined system per supercell, respectively, and $N_C$ is the number of interface carbon atoms in a unit cell. The equilibrium distances between graphene and silicene in Patterns I, II, and III are $d_{Gr-Si}$ = 3.28, 3.57, and 3.41 Å, respectively, with corresponding $E_b$ = −47, −64, and −66 meV per interface C atom. We also checked the possibility of the appearance of Si-C covalent bonds. An initial graphene/silicene structure with covalent Si-C bonds (1.736~2.310 Å) is set and then subject to a fully relaxation without fixing the cell. As shown in the Supplementary Movie S1 we added, silicene and graphene gradually move apart from each other and eventually become stable with a large interlayer distance of 3.49 Å when the maximum residual force is less than 0.005 eV/Å. Therefore, covalent bond cannot be formed between graphene and silicene. In the following study, we focus on the most stable configuration Pattern III.

The geometries of BLG/silicene and TLG/silicene heterostructures are obtained by optimizing the different initial structures by adding one or two more graphene layers on SLG/silicene Pattern III structure. The calculated key data of the most stable structures among the checked heterostructures are presented in Table 1. The binding energy $E_b$ between graphene and silicene is nearly independent of the graphene layer number (−0.066 ~ −0.068 eV per interface C atom), comparable to the exfoliation energy of −0.052 ± 0.005 eV/C atom for graphite.[33] The interlayer distance between graphene and silicene $d_{Gr-Si}$ (3.41 ~ 3.49 Å) is comparable to that of multilayer graphene.

## 2.2. Electronic and Transport Properties of (3×3)Graphene/(2×2)Silicene Heterostructures

The Brillouin zones (BZs) of graphene and silicene and the mini BZ (miBZ) of the heterostructures are shown in Figure 2a. Note that the $K$ points of graphene and silicene ($K_{Gr}$ and $K_{Si}$) are folded to the $\Gamma$ and $K$ points of the miBZ of the heterostructures, respectively (Figure 2b). Consequently, the Dirac cones of graphene and silicene are observed in the vicinity of the $\Gamma$ and $K$ points, respectively, in the miBZ as shown in Figure 2c-2f. The Dirac cones of graphene and silicene are nearly intact when they are stacked on each other except a small band opening ($\leqq$ 0.1 eV) if the inversion symmetry is broken. In $3n \times 3n$ and $\sqrt{3}\,n \times \sqrt{3}\,n$ graphene/silicene/germanene supercells, the band gap opening originates from a coupling between inversion symmetry breaking and the intervalley interaction because the two valleys $K$ and $K'$ are folded to the same $\Gamma$ point and often exhibits a relatively large value.[34-36] The energy range of absence of band hybridization around $E_f$ is 1.3, 1.4, and 1.4 eV for SLG/silicene, BLG/silicene, and TLG/silicene heterostructures, respectively. The wave functions of the SLG/silicene heterostructure at four different ($k$, $E$) points are depicted in Figure 2c. The wave functions near the two Dirac cones are localized in graphene or silicene, and suggest that electron of graphene is difficult to transmit to silicene and vice versa in the non-hybridization region. By contrast, the wave functions of both the degenerate and nondegenerate (see Figure S1) points in the hybridization region are distributed on both graphene and silicene and suggest an easy electron transfer between graphene and silicene. This electron opaque feature can also be understood in terms of the mismatch of energy $E$ and momentum $k$ in graphene and silicene. Correspondingly, a transport gap should exist in a transport process from graphene to silicene in graphene/silicene heterostructures without assistance of phonon.

We simulate a single-gated vertical device based on the (3×3)SLG/(2×2)silicene heterostructure, as shown in Figure 3a and 3b. SLG is *n*-type doped by K atoms in the source region to achieve a high electron sheet density. K atoms are located above the hexagonal ring center of SLG and have a concentration of 1 atom per graphene unit cell. The distance between K atoms and graphene has been optimized. K atoms have been also doped at the metal electrode/WSe$_2$ contacts of WSe$_2$ FETs to improve the electron sheet density and lower the contact resistances.[37] . In a real device, we can use metal electrode that has a low (high) work function such as Ag (Au) to contact graphene surface and *n*-type (*p*-type) dope graphene instead of K atoms.[38] To avoid the doping effect of K atoms on the heterostructures in the channel region, a long buffer zone of pure graphene is set as shown in Figure 3b, whose length is indicated by $L'$. The vertical device has a gate length of $L_g$ = 10 nm, an overlap length between SLG and silicene of $L_o$ = 4 nm, and a buffer zone length of $L'$ = 5 nm. Other overlap region lengths are also investigated, and the main following conclusion is not altered (Figure S2 in supporting information). We adopt periodic boundary condition in the $x$-direction with $a_x$ = 7.5 Å. The transmission probability as a function of energy $E$ and $k_x$ ($T(E$, $k_x$)) at $V_g = 0$ V and $V_{ds} = 0$ V is compared with the folded band structure of SLG/silicene heterostructure in the $k_x$-direction in Figure 3c. This system reveals a prominently large transport gap of 1.3 eV around $E_f$, in good agreement with the energy range of **k**-mismatch (1.3 eV) revealed in Figure 2c.

The $(E, k_x)$ dependent transmission probability $T(E, k_x)$ in the conduction band has a correspondence with the $(E, k_x)$ dependent band hybridization between SLG and silicene. Such a correspondence is not clear in the valence band, and we tentatively attribute this to the valence band warping of SLG due to the K doping effect. To verify this point, we compare $T(E, k_x)$ of the vertical device without K atoms with the folded band structure of SLG/silicene heterostructure in Figure S3. The transport gap remains 1.3 eV, and there is a correspondence between $T(E, k_x)$ and the $(E, k_x)$ dependent band hybridization between SLG and silicene in both the conduction and valence bands. The maxima of $T(E, k_x)$ in Figure 3c and Figure S3a are not always located on the bands. One possible reason is the different basis sets used in the band structure and transmission calculations. Another is that the transmission coefficients $T(E, k_x)$ of a vertical FET are calculated with a finite channel length (10 nm) and can also be affected by the electrodes, while the band comes from a periodical structure.

The total transmission spectra under different gate voltages from −10 to 20 V are given in Figure 3d, where the bias voltage is fixed at $V_{ds} = -0.2$ V. This transmission gap is shifted to the left with the increasing $V_g$ and is moved away from the bias window at $V_g = 20$ V, resulting in a transmission peak around $E_f$. According to Equation 1, the drain current $I_{ds}$ is calculated and then normalized by the overlap area to obtain the current density (Figure 3e). Clear on and off current modulation is achieved by varying the gate voltage. The device shows an on/off current ratio of $4.1 \times 10^7$. If we limit the gate voltage to a more realistic window of −5 ~ 5 V, the on/off current ratio still can reach $7.8 \times 10^4$, which is about 2 orders of magnitude larger than those of the dual-gated BLG and ABC-stacked TLG FETs measured at the room temperature [14, 15] and already sufficient for complementary MOS-like logic. The difference in the on- and off-state is also reflected from the transmission eigenchannel of the device at $E = E_f$ and $k = (1/3, 0)$ as shown in Figure 3f. The transmission eigenvalue of the on-state (defined at $V_g = 20$ V) is 0.758, and most of the incoming wave function is able to reach to the other lead. On the contrary, the transmission eigenvalue of the off-state is $5.521\times10^{-12}$, and the corresponding wave is forbidden to pass through the interface due to the translational symmetry rules. The output characteristic of the vertical FET at $V_g = 0$ V is shown in Figure S4. Although the heterostructure channel is all-metallic, at low $V_{ds}$ ($V_{ds} < 0.4$ V), the current density is strongly suppressed. It implies the existence of a barrier at the SLG and silicene contact because of momentum mismatch. At higher $V_{ds}$ ($V_{ds} > 0.4$ V), the current grows more rapidly.

An all-metallic FET has been proposed on the basis of telescoping pristine double-walled metallic carbon nanotubes (TPDWMCNTs) based on density functional theory (DFT) coupled with nonequilibrium Green's function (NEGF) method, and a high on/off ratio of the conductance is also calculated to be $10^4$. However, the transport gap is nearly zero in TPDWMCNTs, and a finite bias will greatly degrade the current on/off ratio.[39]

The scale effect of the buffer zone is also investigated. As shown in Figure 4a, when the buffer zone length $L'$ decreases from 5, 4 to 3 nm ($L_o = 4$ nm and $L_g$ correspondingly changes from 10, 9 to 8 nm), under $V_{ds} = -0.2$ V the off-current density ($V_g = -5$ V) increases exponentially from 0.2 to 61.6 A/cm$^2$ and the on-current density ($V_g = 5$ V) exhibits a relatively weak $L'$-dependence. Therefore the on/off current ratio decreases greatly from $7.8\times10^4$ to 113 with the decrease of $L'$ (Figure 4b), implying the important role of a long buffer region in preventing affection from K atoms in the channel region. In light of the device is asymmetric (the source is K-doped graphene but the drain is pure silicene), we examine the transport dependence on the direction of the bias. As shown in Figure 4a, the off currents under $V_{ds} = -0.2$ and 0.2 V are almost the same in a vertical FET, whereas the on current at $V_{ds} = -0.2$ V is generally slightly larger than that at $V_{ds} = 0.2$ V, showing a weak rectification effect. Thus the on/off ratio is generally larger under $V_{ds} = -0.2$ V than that under $V_{ds} = 0.2$ V (Figure 4b). Because momentum mismatch near $E_f$ also exists in multilayer graphene/silicene heterostructures (Figure 2), a transport gap of 1.3 eV has been observed in the BLG/silicene vertical FET (Figure S5).

## 2.3. (8×8)SLG/(5×5)Silicene Supercell

To examine whether the transport gap remains when both the $K_{Gr}$ and $K_{Si}$ points are folded to the K points of the miBZ, we construct a larger (8×8)SLG/(5×5)silicene supercell with $a_S =$ 19.49 Å. In this case, the constant deviation is 1.0% and 0.8% in graphene and silicene, respectively. As shown in Figure 5a, the band structure has a similar profile with that of the stacked (8×8)SLG/(5×5)silicene/(8×8)SLG.[40] The Dirac cones of SLG and silicene remain intact and located above and below $E_f$, respectively, resulting in a roughly circular intersection at $E_f$ (inset in Figure 5b). Band component analysis indicates that the π states of SLG and silicene don't hybridize with each other near $E_f$ even at the circular intersection region (in terms of the wave function analyses at the cross point in the *M-K* high symmetry line, as shown in Figure 5c). Band hybridization only occurs when $|E - E_f| > 0.6$ eV. Then we construct a vertical FET based on the (8×8)SLG/(5×5)silicene heterostructure. A transport gap of 1.2 eV is observed around $E_f$ (Figure 5b), which is consistent with the non-hybridization energy region in the band structure. By applying a gate voltage, a large current modulation also can be expected.

### 2.4. Rotational Situations

There is an orientation degree of freedom in SLG/silicene heterostructures. Taking the (3×3)SLG/(2×2)silicene heterostructure as an example, when silicene is rotated by an angle $\theta$ relative to SLG, the Dirac cones of both SLG and silicene will survive in an incommensurate rotation because the Dirac cones of SLG and silicene do not coincide after rotation. A transport gap is expected in the corresponding vertical FETs. However, this is difficult to justify from a simulation due to the loss of commensurate condition. A commensurate rotation occurs when $T_{mn} = m a^1_{S/Gr} + n a^2_{S/Gr} = m' a^1_{S/Si} + n' a^2_{S/Si} = T'_{m'n'}$, where $a^1_{S/Gr(Si)}$ and $a^2_{S/Gr(Si)}$ are the two unit vectors of the (3×3)SLG or (2×2)silicene supercell, with $|a_{S/Gr(Si)}| = a_S = 7.5$ Å. The corresponding rotation angle is discrete $\theta_{mn} = \arg[\frac{m e^{-i\pi/6} + n e^{i\pi/6}}{n e^{-i\pi/6} + m e^{i\pi/6}}]$, indexed by the two integers *m* and *n*. The lattice constant of the commensuration supercell is $L = |T_{mn}| = a_S \sqrt{m^2 + n^2 + mn} = \frac{|m-n|}{2\sin(\theta_{mn}/2)} a_S$. The Dirac cones of SLG and silicene are originally located at the Γ and *K* points of the miBZ. After a rotation of $\theta_{mn}$, the Dirac cone of SLG remains located at the Γ point in the reduced BZ ($\Gamma_r$ in the rBZ) of the twisted SLG/silicene heterostructures, while that of silicene is folded to the rBZ corners ($K_r$ and $K'_r$), as shown in Figure 6a, in which $\theta_{12} = 21.8°$. The Dirac cones of the 21.8° twisted (3× 3)SLG/(2×2)silicene heterostructure indeed survive (Figure 6b), and the energy range of absence of band hybridization is about 0.8 eV around $E_f$, which is two-thirds of that without relative rotation. The narrowing of the non-hybridization region is attributed to the fact that the Dirac cone of SLG get closer to that of silicene in the ($E$, $\bm{k}$)-space (the distance of the two Dirac cones in $\bm{k}$ is decreased by a factor of √7). Consistently, a smaller transport gap of 0.6 eV is observed in the corresponding vertical FET under $V_g$ = 0 V and $V_{ds}$ = 0 V, as shown in Figure 6c.

**2.5. A Possible Method to Realize the Graphene/Silicene Heterostructure FET**

Herein we propose a possible procedure to realize the graphene/silicene heterostructure FET. Silicene has been grown on Ag,[3-5] ZrB$_2$,[6] Ir,[7] and MoS$_2$ [8] substrates. Unfortunately, the Dirac cone of silicene is always destroyed as a result of either a strong interaction (Ag, Ir, and ZrB$_2$ cases) or a serious lattice mismatch (MoS$_2$ case).[8, 41-43] It appears that a proper substrate to grow silicene without destroying its Dirac cone should have a weak interaction and matched lattice constant. The interactions between silicene and group III monochalcogenide (G3MC) GaS/GaSe/GaTe are weak van de Waals force (*e.g.* $E_b$ = 0.126 eV per Si)[44] and comparable with that ($E_b$ = 0.2 eV per Si ) between silicene and MoS$_2$.[45] But the lattice mismatches (3.0% ~ 7.4%) between silicene ($a_{Si}$ = 3.866 Å) and GaS/GaSe/GaTe ($a$ = 3.580 ~ 4.100 Å) are smaller than that (17.7%) between silicene and MoS$_2$ ($a$ = 3.180 Å). The Dirac cone of silicene is kept on GaS/GaSe/GaTe substrate based on our and other DFT calculations.[45] Therefore, GaS/GaSe/GaTe appears to be a proper substrate to grow silicene. It might be interesting to check whether this scheme can be done experimentally. Graphene is then transferred on the top of silicene if silicene could be grown on GaS/GaSe/GaTe substrate. The band structure of the (3×3)SLG/(2×2)silicene on the (2×2)GaS substrate is shown in Figure 7. The Dirac cone of silicene is preserved on GaS substrate although it is destroyed on MoS$_2$ substrate. The band components of the (3×3)SLG/(2×2)silicene are intact compared with those without GaS nanosheet shown in Figure 2c and the energy range of absence of band hybridization remains about 1.3 eV.

**2.6. Other Stacked Dirac Materials**

A (5×5)SLG/(3×3)germanene heterostructure is also constructed, with $a_S$ = 12.25 Å and lattice mismatch of 0.4% and 0.5% in SLG and germanene, respectively. As shown in Figure 8a and 8b, the energy range of absence of band hybridization is 1.1 eV around $E_f$, and the transport gap is 0.9 eV in its vertical FET. We expect that in any vertical heterostructure consisting of two weakly interacting Dirac materials, the Dirac cones will survive as long as these Dirac cones do not coincide. As a result, a transport gap and a high current on/off ratio can be expected if no phonon takes parts in the electron transport.

Suppression of transmission by mismatch of $k_\parallel$ (transverse momentum parallel to the grain boundary) has been calculated in polycrystalline SLG with specific grain boundary structures through *ab initio* quantum transport calculations, exhibiting a transport gap of 1.0 eV and an on/off current ratios above $10^3$.[46] It is an interesting question: Does a transport gap exist in a bilayer homogenous Dirac material? In a twisted BLG with an incommensurate rotation, electron transmission from the Dirac cones between different layers should be forbidden because the Dirac cones of the two graphene layers are separated from each other (Figure S6). A transport gap of $E_g^{trans} = \eta v_f \Delta K = \frac{4}{3a_{Gr}} h v_f \sin(\frac{\theta}{2})$ is expected in the corresponding devices, where $v_f$ is the Fermi velocity (~$10^6$ m/s), $\Delta K$ is the distance between the $K$ points of the two layers in $\boldsymbol{k}$-space, $a_{Gr}$ = 2.46 Å is the lattice constant of graphene, and $\theta$ is the rotation angle. For an incommensurate rotation when 2°<$\theta$<58°, a $E_g^{trans}$ > 0.4 eV will be obtained. Unfortunately, a device simulation is unfeasible for an incommensurate system. The two layers become commensurate only when $\boldsymbol{T}_{mn} = m\boldsymbol{a}_{Gr1}^1 + n\boldsymbol{a}_{Gr1}^2 = m'\boldsymbol{a}_{Gr2}^1 + n'\boldsymbol{a}_{Gr2}^2 = \boldsymbol{T}'_{m'n'}$, where $\boldsymbol{a}_{Grl}^1$ and $\boldsymbol{a}_{Grl}^2$ are the two lattice vectors of the $l$ layer ($l$ = 1 and 2). The corresponding rotation angle is $\theta_{mn} = \arg[\frac{me^{-i\pi/6} + ne^{i\pi/6}}{ne^{-i\pi/6} + me^{i\pi/6}}]$. No matter even or odd the sublattice exchange parity is, the $\pi$ states of the two layers strongly hybridize with each other due to the interlayer intravalley or intervalley coupling.[47] Indeed, there is no transport gap in a vertical commensurate BLG FET (Figure S7). Even so, a pseudogap appears when the sublattice exchange is even (Figure S7, $\theta_{14}$ = 38.2°).

In order to confirm that there is no band hybridization when two Dirac cones of BLG don't coincide, we rotate one layer by 30° and impose a biaxial strain of $\varepsilon$ = 15.5% on it. Such a strain is practical since the elastic strain of SLG has been measured experimentally to be as high as 25%.[48] We obtain a (√3×√3)SLG/(2×2)SLG supercell with a translation invariation. A band gap of 0.6 eV is opened in the stressed (√3×√3)SLG due to a coupling between inversion symmetry breaking and the intervalley interaction when both the $K_{Gr}$ and $K_{Gr}'$ are folded into the Γ point.[17] As expected, no band hybridization is observed in a large energy range of 2.8 eV around $E_f$ in the band structure of the (√3×√3)SLG/(2×2)SLG and a transport gap of 2.1 rather than 0.6 eV appears in its vertical FET simulation (Figure 8c and 8d).

It should be pointed out that in our calculation phonon is not taken into account at all. At a finite temperature, there is a certain possibility of electron transport from one Dirac cone of one Dirac material to that of the other Dirac materiel due to the existence of phonon, which can provide the required momentum. Therefore, the performance of the vertical FETs composed of stacked Dirac materials should be degraded to a certain extent.

2D materials, with every atom on the surface, show a molecule-like sensitivity to its surroundings. Therefore, it is important to improve the environmental resistance of our devices. hBN films have been proved to be chemically stable high-temperature coatings for graphene,[49, 50] silicene,[11] and germanene[51] without degrading their electronic properties. Therefore, it's an effective method to protect our devices from the oxide substrate and environment by sandwiching the stacked Dirac materials with hBN films.

## 3. Conclusions

Like a conventional metal, the electron motion in Dirac materials such as graphene, silicene, and gemanene, is also difficult to control by a gate voltage. We find that the Dirac cones belong to different layers are robust against the weak van der Waals interlayer interaction if they are well separated in the reciprocal space (such as graphene/silicene heterostructure and twisted bilayer graphene), leading to a forbidden electron transfer from one Dirac cone of one layer to that in other layer without assistance of phonon based on DFT calculations. Dirac material vertical structures can have an extraordinary on/off current ratio of over $10^4$ due to a large transport gap as long as the Dirac cones of different layers do not coincide based on subsequent *ab initio* quantum transport simulations. Therefore, all-metal field effect transistors with high switching ability are expected to be realized in Dirac materials. Very recently, by alignment of the crystallographic orientation of two graphene layers in a graphene/hBN/graphene heterostructure, resonant tunneling with both electron energy and momentum conservation and negative differential conductance are achieved with stable oscillations in the megahertz frequency range.[52] Experimental work aimed at observing a large current switching ratio in twisted bilayer graphene is under way.

## 4. Computational Details

We carry out the geometry optimizations by employing the CASTEP package [53] with the ultrasoft pseudopotential [54] and plane-wave basis set. The cut-off energy is 350 eV. To take the dispersion interaction between the two Dirac materials into account, a DFT-D semiempirical dispersion-correction approach is adopted with the Tkatchenko-Scheffler (TS) scheme,[55] which once predicts the binding energy and the interfacial distance of graphene on metals in good agreement with the experimental values.[38] A vacuum slab more than 15 Å is set to avoid spurious interaction between periodic images. The maximum residual force is less than 0.005 eV/Å. The electronic structures are calculated with the projector-augmented wave (PAW) pseudopotential [56, 57] and plane-wave basis set with a cut-off energy of 500 eV implemented in the Vienna *ab initio* simulation package (VASP) in order to analyze the band components.[58-61] The Monkhorst-Pack [62] $k$-point mesh is sampled with 6×6 and 12×12 in the BZ during the relaxation and electronic calculation periods, respectively. The dipole corrections are included in both the relaxation and electronic calculations. The zero-field geometry and band structures generated from CASTEP and VASP packages coincide well.

A gated two-probe model is used to simulate the transport properties of graphene/silicene heterostructures. We use pristine or doped SLG as source and pristine silicene as drain. The test shows that the on/off ratio is quite poor without K doping. The thickness of the dielectric regions is $d_i$ = 10 Å, and the dielectric constant is taken as $\varepsilon$ = 3.9, which models $SiO_2$. Bottom gate is used. Gate voltage is applied in $y$ direction where Neumann boundary condition is used. Periodic boundary is used in $x$ direction, and continuous boundary is used in $z$ direction. Electron static potential is obtained by solving Poisson equation using multi-grid method. Transport properties are calculated by using fully self-consistent NEGF method and DFT, which are implemented in ATK 11.2 package.[63-65] Single-$\zeta$ (SZ) basis set is used. The real-space mesh cutoff is 150 Ry, and the temperature is set at 300 K. The entire treatment of transport in this context is ballistic and elastic, *i.e.* there is no scattering by phonons. The electronic structures of electrodes and central region are calculated with a Monkhorst–Pack [62] 50×1×50 and 50×1×1 *k*-point grid, respectively. The current is calculated using the Landauer-Büttiker formula:[66]

$$I(V_g, V_{ds}) = \frac{2e}{h} \int_{-\infty}^{+\infty} \{T_{V_g}(E, V_{ds})[f_L(E-\mu_L) - f_R(E-\mu_R)]\}dE \quad (1)$$

where $T_{V_g}(E, V_{ds})$ is the transmission probability at a given gate voltage $V_g$ and bias voltage $V_{ds}$, $f_{L/R}$ the Fermi-Dirac distribution function for the left (L)/right (R) electrode, and $\mu_L/\mu_R$ the electrochemical potential of the L/R electrode. Generalized gradient approximation (GGA) of Perdew–Burke–Ernzerhof (PBE) form [67] to the exchange-correlation functional is used throughout this paper.

**Supporting Information**

Supporting Information is available from the Wiley Online Library or from the author.

**Acknowledgments**

This work was supported by the National Natural Science Foundation of China (Nos. 11274016, 51072007, 91021017, 11047018, and 60890193), the National Basic Research Program of China (Nos. 2013CB932604 and 2012CB619304), Fundamental Research Funds for the Central Universities, and National Foundation for Fostering Talents of Basic Science (No. J1030310/No. J1103205). Y. Wang also acknowledges the financial support from the China Scholarship Council.

**Table 1.** $E_b$ is the binding energy per interface C atom of graphene/silicene heterostructures; $d_{\text{Gr-Si}}$ is the vertical distance between graphene and silicene, and $d_{\text{Gr-Gr}}$ the interlayer vertical distance of graphene. $\Delta$ is the buckling distance of silicene as indicated in Figure 1. $\Delta E_f^{\text{Gr(Si)}}$ and $E_g^{\text{Gr(Si)}}$ are the Fermi level shifts and opened band gaps in graphene (silicene).

| material | $E_b$ (eV) | $d_{\text{Gr-Si}}$ (Å) | $d_{\text{Gr-Gr}}$ (Å) | $\Delta$(Å) | $\Delta E_f^{\text{Gr}}$ (eV) | $\Delta E_f^{\text{Si}}$ (eV) | $E_g^{\text{Gr}}$ (eV) | $E_g^{\text{Si}}$ (eV) |
|---|---|---|---|---|---|---|---|---|
| SLG/silicene | −0.066 | 3.41 | -- | 0.64 | −0.32 | 0.26 | 0 | 0.026 |
| BLG/silicene | −0.067 | 3.35 | 3.21 | 0.60 | −0.18 | 0.21 | 0.101 | 0.044 |
| TLG(ABA)/silicene | −0.068 | 3.49 | 3.28[a]/3.51[b] | 0.66 | −0.08 | 0.11 | 0 | 0.043 |
| TLG(ABC)/silicene | −0.067 | 3.49 | 3.27[a]/3.53[b] | 0.66 | −0.13 | 0.17 | 0.103 | 0.044 |

[a] The vertical distance between the nearest and next nearest graphene layer with respect to silicene.

[b] The vertical distance between the next nearest and farthest graphene layer with respect to silicene.

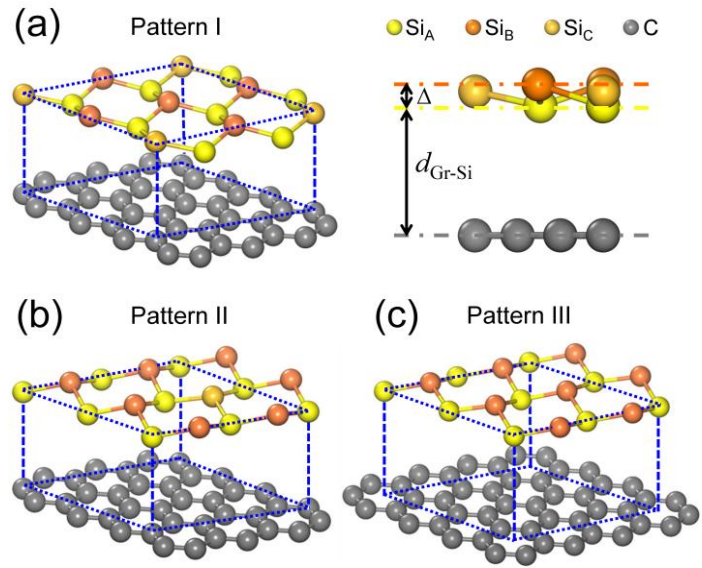

**Figure 1.** (3×3)SLG/(2×2)silicene heterostructure supercells. Side views of the stacking Pattern I-III are shown in a)-c), respectively. Height profile of Pattern I is provided on the right of panel of a). Different types of Si atoms $Si_A$, $Si_B$, $Si_C$ and carbon atoms are denoted with golden yellow, reddish orange, yellow, and gray balls, respectively. The dashed blue lines represent the boundary of the supercell.

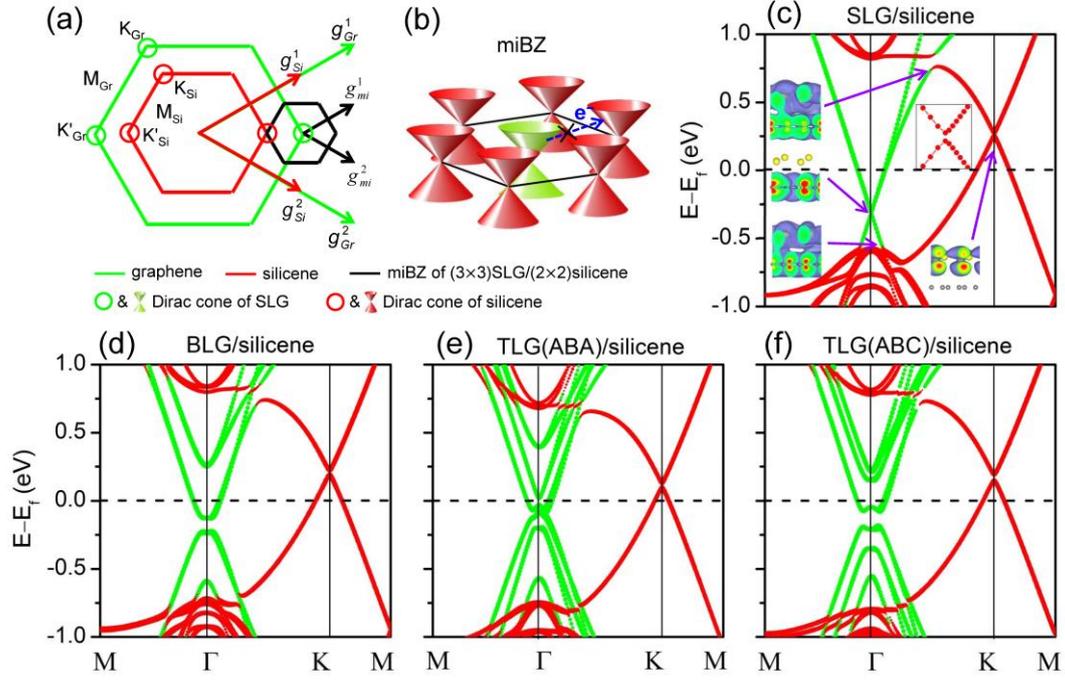

**Figure 2.** a) Brillouin zones (BZs) of the silicene and graphene lattices drawn to scale. Mini BZ (miBZ) of the (3×3)SLG/(2×2)silicene heterostructure is indicated around the $K$ point of graphene ($K_{Gr}$). b) Dirac cones of graphene (green) and silicene (red) in the miBZ. c)-f) Band structures of the SLG, BLG, ABA- and ABC-stacked TLG/silicene heterostructures. Silicene-dominated bands (red) are plotted against the graphene projected bands (green). The wave functions at four ($k$, $E$) points are represented in the insets of c), with an isovalue of 0.002 $e$/ Å$^3$.

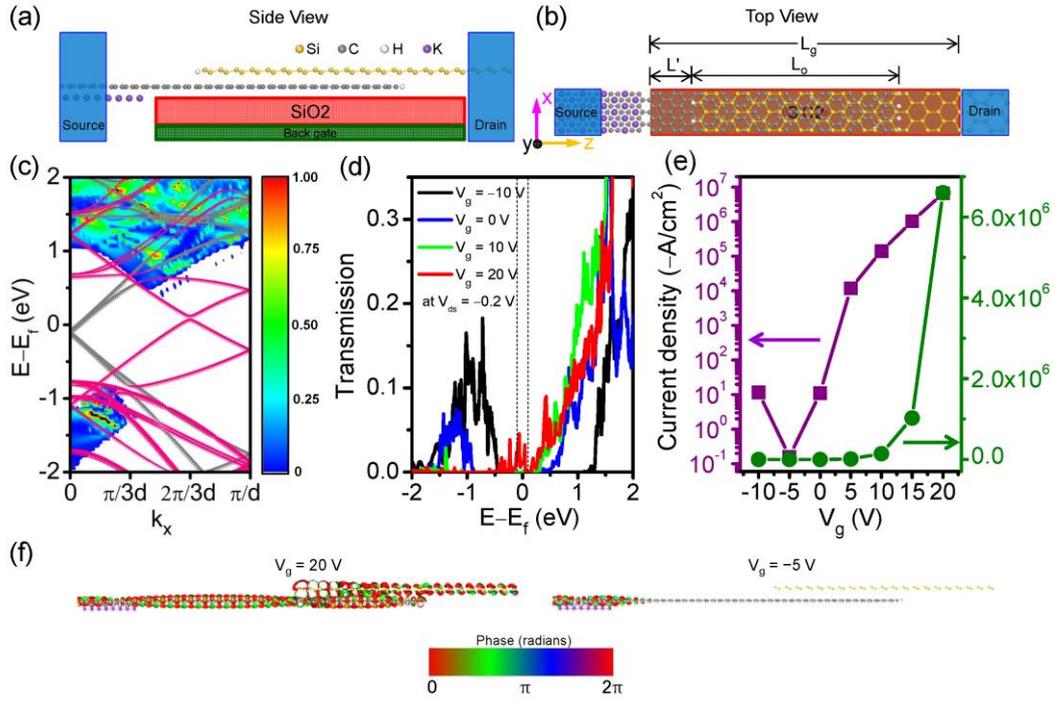

**Figure 3.** Single-gated (3×3)SLG/(2×2)silicene heterostructure vertical FET: a) Side and b) top views of the schematic model. c) Comparison between the ($E$, $k_x$) dependent transmission probability under $V_g$ = 0 V and $V_{ds}$ = 0 V and the folded band structure of the SLG/silicene heterostructure in the $k_x$-direction (gray and pink lines denote SLG and silicene components, respectively). The color scale is shown on the right. d) Transmission spectra under different gate voltages. e) Transfer characteristics at $V_{ds}$ = −0.2 V. f) Transmission eigenstates at $E = E_f$ and at $k$ = (1/3, 0) under $V_{ds}$ = −0.2 V for $V_g$ = 20 and −5 V, respectively. The isovalue is 0.2 a.u..

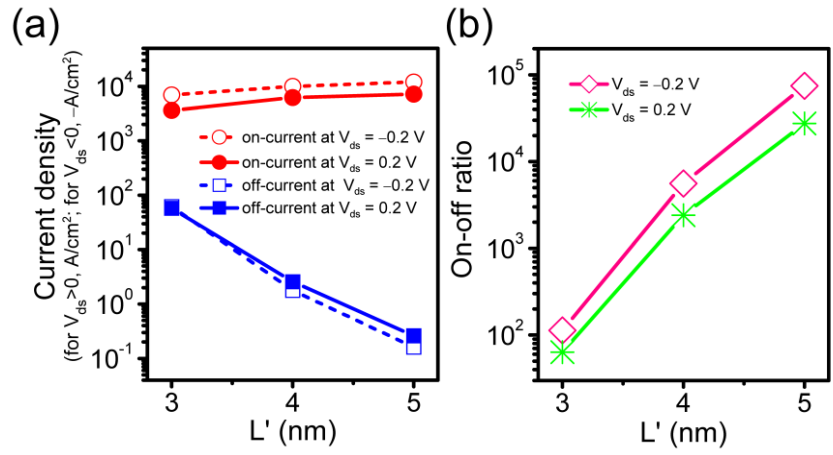

**Figure 4.** a) On- and off-current densitys and b) on/off current ratio under $V_{ds} = -0.2$ and 0.2 V in the local-gated (3×3)SLG/(2×2)silicene heterostructure device with $L'$ = 3, 4, and 5 nm, respectively. The gate voltage range is −5 ~ 5 V.

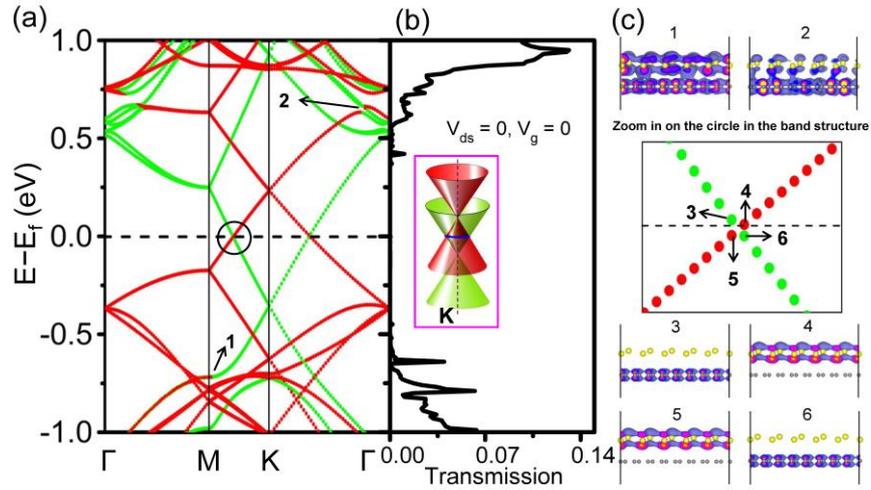

**Figure 5.** a) Band structure of the (8×8)SLG/(5×5)silicene heterostructure. b) Transmission spectrum of this heterostructure vertical FET with $L_g$ = 10 nm and $L_o$ = 4 nm under $V_g$ = 0 V and $V_{ds}$ = 0 V. Inset: Dirac cones of graphene (green) and silicene (red) at the *K* point. Their intersection is donated in blue. c) Wave functions at six (*k*, *E*) points indicated in panel a), with an isovalue of 0.003 *e*/ Å$^3$.

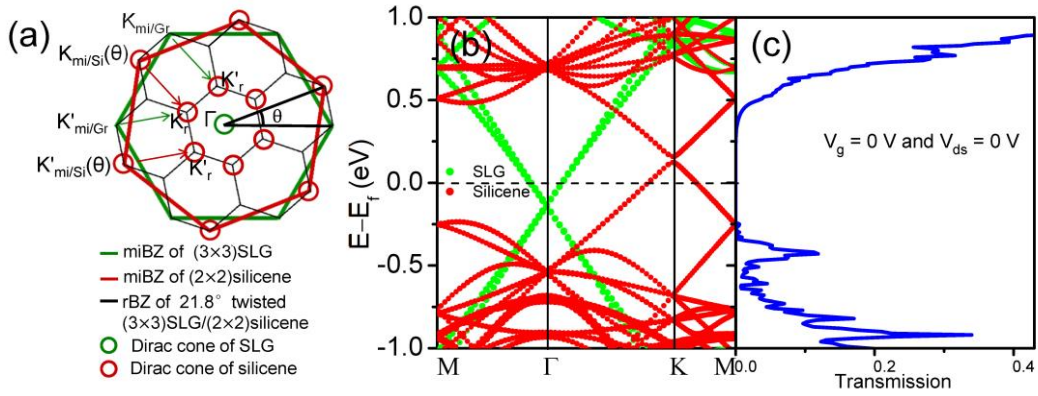

**Figure 6**. a) miBZs of (3×3)SLG (green hexagon) and (2×2)silicene (red hexagon) with a relative rotation $\theta_{12} = 21.8°$ and reduced BZ (rBZ) of the rotated (3×3)SLG/(2×2)silicene heterostructure (small black hexagon). Green and red circles represent the locations of Dirac cones of SLG and silicene, respectively. b) Band structure of the 21.8° rotated (3×3)SLG/(2×2)silicene heterostructure. c) Transmission spectrum of the corresponding vertical FET under $V_g = 0$ V and $V_{ds} = 0$ V.

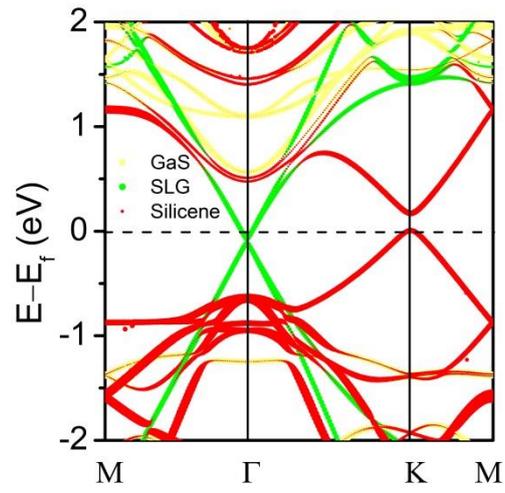

**Figure 7.** Band structure of (3×3)SLG/(2×2)silicene on the (2×2)GaS substrate.

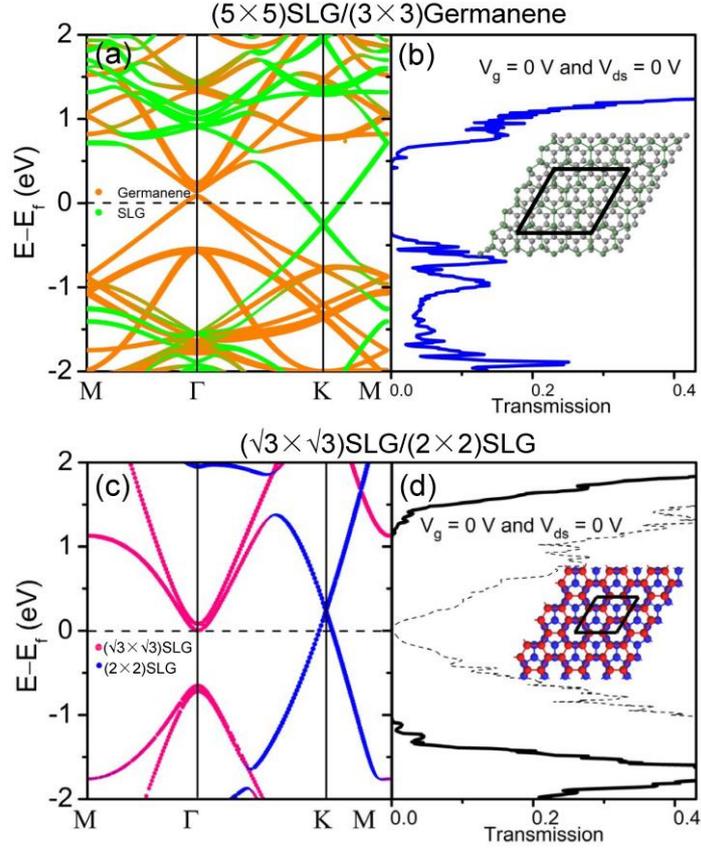

**Figure 8.** Band structures and the corresponding transmission spectra of the [a), b)] (5×5)SLG/(3×3)germanene and [b), c)] (√3×√3)SLG/(2×2)SLG structures (a biaxial strain $\varepsilon$ = 15.5% is imposed in the (√3×√3)SLG). The transmission spectrum of the untwisted BLG (dashed line) is provided in d) for comparison. Inset in b): the (5×5)SLG/(3×3)germanene supercell. Inset in d): the (√3×√3)SLG/(2×2)SLG supercell.

TOC:

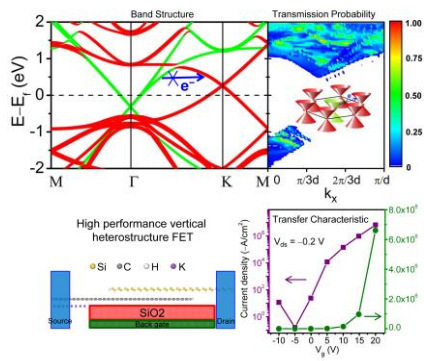

**Electron transport from one Dirac material to the other near $E_f$ is forbidden** by momentum mismatch if the two Dirac cones of different layers are well separated. All-metallic field effect transistor can be designed out of Dirac materials with a large transport gap and a high on/off current ratio of over $10^4$ based on *ab initio* quantum transport simulations.